\documentclass[journal]{IEEEtran}
\usepackage{setspace}

\usepackage{amssymb}
\usepackage{amsmath}    
\usepackage{graphicx}
\usepackage{epsfig}
\usepackage{epsf}
\usepackage{psfrag}
\usepackage{url}
\usepackage{algorithm}
\usepackage{algorithmic}
\usepackage{booktabs}

\newcommand{\qed}{\hfill \ensuremath{\Box}}

\usepackage{cite}

\ifCLASSINFOpdf
\else
\fi
\usepackage{graphicx}

\newtheorem{thm}{Theorem}[section]

\newtheorem{prop}[thm]{Proposition}

\begin{document}
%
\title{{{ A Distributed Sequential Algorithm for Collaborative Intrusion Detection Networks}}}
\IEEEoverridecommandlockouts
\author{\IEEEauthorblockN{Quanyan Zhu}
\IEEEauthorblockA{Coordinated Science Laboratory\\Department of Electrical and Computer Engineering\\
University of Illinois at Urbana Champaign\\
USA 61820\\
Email: zhu31@uiuc.edu}
\and
\IEEEauthorblockN{Carol Fung}
\IEEEauthorblockA{School of Computer Science\\
University of Waterloo\\
Waterloo, Ontario \\
Email: j22fung@cs.uwaterloo.ca}}

\author{\IEEEauthorblockN{
    Quanyan~Zhu$^\textbf{1}$,
    Carol~J.~Fung$^\textbf{2}$,
    Raouf~Boutaba$^\textbf{2}$, and
    Tamer~Ba\c{s}ar$^\textbf{1}$
}
    \IEEEauthorblockA{
         $^\textbf{1}$Department of Electrical and Computer Engineering,\\ University of Illinois at Urbana-Champaign, USA.
        \{zhu31, basar1\}@illinois.edu\\
                        $^\textbf{2}$David R. Cheriton School of Computer Science,\\ University of Waterloo, Ontario, Canada,
        \{j22fung, rboutaba\}@uwaterloo.ca
    }

}

\author{Quanyan Zhu, Carol J. Fung, Raouf Boutaba and Tamer Ba\c{s}ar \vspace{-9mm}
\thanks{Quanyan Zhu  and Tamer Ba\c{s}ar are with the Department of Electrical and Computer Engineering, University of Illinois at Urbana Champaign, 61801 (email:\{zhu31,basar1\}@illinois.edu). Carol J. Fung and Raouf Boutaba are with David R. Cheriton School of Computer Science, University of Waterloo, Ontario, Canada (email:    \{j22fung, rboutaba\}@uwaterloo.ca).}
\thanks{The work of the authors from University of Illinois was in part supported by a grant from Boeing through the Information Trust Institute. The work of the authors from the University of Waterloo is supported by the Natural Science and Engineering Research Council of Canada under its strategic program and in part by WCU (World Class University) program through the Korea Science and Engineering Foundation funded by the Ministry of Education, Science and Technology (Project No. R31-2008-000-10100-0).}
}
\maketitle

\begin{abstract}
Collaborative intrusion detection networks are often used to gain  better detection accuracy and cost efficiency as compared to a single host-based intrusion detection system (IDS). Through cooperation, it is possible for a local IDS to detect new attacks that may be known to other experienced acquaintances. In this paper, we present a sequential hypothesis testing method for feedback aggregation for each individual IDS in the network. Our simulation results corroborate our theoretical results and demonstrate the properties of cost efficiency and accuracy compared to other heuristic methods. The analytical result on the lower-bound of the average number of acquaintances for consultation is essential for the design and configuration of IDSs in a collaborative environment.
\end{abstract}


%
\IEEEpeerreviewmaketitle

\section{Introduction}

As computer systems become increasingly complex, the accompanied potential threats also grow to be more sophisticated.
Intrusion detection is the process of monitoring and identifying attempted unauthorized system access or manipulation. It is one of the most important tools for a network administrator to detect security breaches along with firewalls. 

An IDS can be categorized as either host-based or network-based. A host-based IDS (HIDS) is intended primarily to monitor a host, which can be a server, workstation, or any networked device, whereas a network-based IDS (NIDS) is used to protect a group of computer hosts by capturing and analyzing network packets. Even though these two types of IDSs are commonly employed in an enterprise network, they do not adequately leverage the possible information exchange between IDSs. The exchange of alert data or decisions between administrative domains can effectively supplement the knowledge gained by a single local IDS. In a collaborative environment, an IDS can learn the global state of network attack patterns from its peers. By augmenting the information gathered from across the network, an IDS can have a more precise picture of an attacker's behavior and hence increase its accuracy and efficiency of detection. 

Collaborative intrusion detection networks (CIDNs) have distinct features from some other types of social networks such as P2P network and E-commerce network, where the collaboration is one-time or short-term pattern. The collaboration in IDN is usually long-term based. Unlike other social networks, communication in CIDNs is often of ``low-cost'', which leads to the possibility of using {\itshape test messages} (a communication overhead  generated on purpose to test the reliability of the collaborators). 

\begin{figure}[t]
\centering
\includegraphics[scale=0.32]{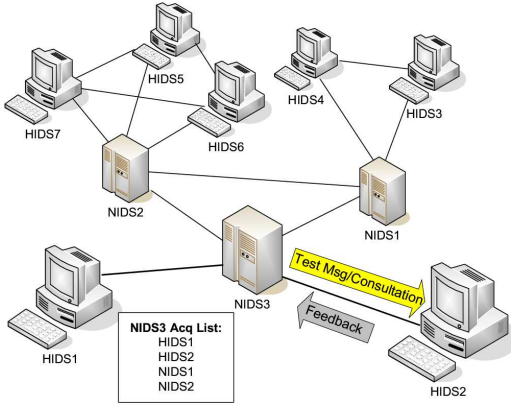}
\caption{A Collaborative Intrusion Detection Network} \label{fig:LNIDS}
\end{figure}

Based on the aforementioned properties, we design a CIDN which utilizes {\itshape test messages} to learn the reliability of others and {\itshape consultation requests} to seek diagnosis from collaborators. The architecture design is shown in Figure \ref{fig:LNIDS}, where NIDSs and HIDSs are connected into a collaboration network. Each IDS maintains a list of {\itshape acquaintances} (collaborators) and test messages are sent to acquaintances periodically to update its belief on peer reliability. When an IDS receives intrusion alerts and lacks confidence to determine the nature of the alerted source, alert messages are sent to its acquaintances for evaluation. An acquaintance IDS analyzes the received intrusion information and replies with a {\itshape feedback} of  positive/negative diagnosis. The ambivalent IDS collects feedback from its acquaintances and decides whether an alarm should be raised or not to the administrator. If an alarm is raised, the suspicious intrusion flow will be suspended and the system administrator investigates the intrusion immediately. 

In this paper, we design an efficient distributed sequential algorithm for IDSs to make decision based on the feedback from its collaborators. We investigate four possible outcomes of a decision: {\itshape false positive} (FP), {\itshape false negative} (FN), {\itshape true positive} (TP), and {\itshape true negative} (TN). 
Each outcome is associated with a cost. Our proposed sequential hypothesis testing based feedback aggregation provides improved cost efficiency as compared to other heuristic methods, such as the simple average model~\cite{resnick2006value} and the weighted average model~\cite{Duma06, Fung2009}. In addition, the algorithm reduces the communication overhead as it aggregates feedback until a predefined FP and TP goal is reached. Our analytical model effectively estimates the number of acquaintances needed for an IDS to reach its predefined intrusion detection goal. Such result is crucial to the design of  an IDS acquaintance list in CIDN.

The remainder of this paper is organized as follows. In Section II, we review some existing CIDNs in the literature and IDS
feedback aggregation techniques. The problem formulation is in Section III, where we use hypothesis testing to minimize the cost of decisions and sequential hypothesis testing to form consultation termination policy for predefined goals. In Section IV, we use a simulation approach to evaluate the effectiveness of our aggregation system and validate the analytical model. Section V concludes the paper and identifies directions for future research.

\section{Related Work} \label{relate}
Many CIDNs were proposed in the literature, such as Indra~\cite{Janakiraman03}, DOMINO~\cite{Yegneswaran04globalintrusion}, and NetShield~\cite{cai2005collaborative}. However, these works did not address the problem that the system might be degraded by some compromised insiders who are dishonest or malicious. 

Simple majority voting~\cite{ghosh2004agent} and trust management are commonly used to detect malicious insiders in CIDNs. Existing trust management models for CIDN are either linear as in \cite{Duma06},~\cite{Fung2008}, or Bayesian model as in ~\cite{Fung2009}. They are based on heuristic where the feedback aggregation is either a simple average ~\cite{resnick2006value} or a weighted average~\cite{Fung2009}. Moreover, no decision cost is considered in these models. In this paper, we use a sequential hypothesis testing model aiming at finding cost-minimizing decisions based on collected feedback.
~
Existing work that applies hypothesis testing for intrusion detection includes\cite{tsitsiklis1993decentralized} and \cite{nguyen2008dba}, where a central data fusion center is used to aggregate results from distributed sensors in a local area network. However, their methodologies are limited to the context that all participants need to engage in every detection case. While in our context, IDSs may not be involved in all intrusions detection and the collected responses may be from different groups of IDSs each time.

\section{Problem Formulation}

In this section, we formulate the feedback aggregation as a sequential hypothesis testing problem. Consider a set of $N$ nodes, $\mathcal{N}$, connected in a network, which can be represented by a graph $\mathcal{G}=(\mathcal{N}, \mathcal{E})$. The set $\mathcal{E}$ contains the undirected links between nodes, indicating the acquaintances of IDSs in the network. 

Let $Y_i, i\in\mathcal{N},$ be a random variable denoting the decision of IDS $i$ observed by its peer IDSs on its acquaintance list $\mathcal{N}_i$. The random variable $Y_i$ takes values in $\mathcal{Y}_i=[0, 1]$. In the intrusion detection setting, $Y_i=0$ says that IDS $i$ decides that there is no intrusion while $Y_i=1$ means that IDS $i$ raises an alarm of possible detection of intrusion. Each IDS makes its decision based upon its own experience of the previous attacks and its own sophistication of detection. We let $p_i$ as the probability mass function defined on $\mathcal{Y}_i$ such that $p_i(Y_i=0)$ and $p_i(Y_i=1)$ denote the probability of no intrusion and the probability of intrusion from $i$, respectively.

We let $\mathbf{Y}^i:=[Y_j]_{j\in\mathcal{N}_j}\in \mathcal{Y}^i:=\prod_{j\in\mathcal{N}_i}\mathcal{Y}_i$ be an observation vector of an IDS $i$ that contains the feedback from its peers in the acquaintance list. Each IDS has two hypotheses $H_0$ and $H_1$. $H_0$ hypothesizes that no intrusion is detected whereas $H_1$ forwards a hypothesis that intrusion is detected and alarm needs to be raised. Note that we intentionally drop the superscript $i$ because we assume that each IDS attempts to make the same decision. Denote by $\pi_0^i, \pi_1^i$ the apriori probabilities on each hypothesis such that $\pi_0^i=\mathbb{P}[H_0], \pi_1^i=\mathbb{P}[H_1]$ and $\pi_0^i+\pi_1^i=1$, for all $i\in\mathcal{N}$.
The conditional probability $p^i(\mathbf{Y}^i=\mathbf{y}^i|H_l), l=1, 2$ denotes the probability of a complete feedback being $\mathbf{y}^i\in\prod_{j\in\mathcal{N}_i}\mathcal{Y}_j$ given the hypothesis. Assuming peers make decisions independently (this is reasonable if acquaintances are appropriately selected), we can rewrite the conditional probability as
{\small \begin{equation}\label{condProb}
p^i(\mathbf{Y}^i=\mathbf{y}^i|H_l)=\prod_{j\in\mathcal{N}_i}
p_j(Y_j=y_j|H_l), i\in\mathcal{N}, l=0,1.
\end{equation}}
A hypothesis testing problem is one of finding a decision function $\delta^i(\mathbf{Y}^i): \mathcal{Y}^i\rightarrow \{0,1\}$ to partition the observation space $\mathcal{Y}^i$ into two disjoint sets $\mathcal{Y}_0^i$ and $\mathcal{Y}_1^i$, where $\mathcal{Y}_0^i=\{\mathbf{y}^i: \delta^i(\mathbf{y}^i)=0\}$, and
$\mathcal{Y}_1^i=\{\mathbf{y}^i:\delta^i(\mathbf{y}^i)=1\}$.
 
 To find an optimal decision function according to some criterion, we introduce the cost function $C^i_{ll'}, l,l'=0, 1$, which represents IDS $i$'s cost of deciding that $H_l$ is true when $H_{l'}$ holds. More specifically, $C^i_{01}$ is the cost associated with a missed intrusion or attack and $C^i_{10}$ refers to the cost of false alarm, while $C^i_{00}, C^i_{11}$ are the incurred costs when the decision meets the true situation. In several situations, it can be shown that decision functions can be picked as function of the likelihood ratio given by $L^i(\mathbf{y}^i)=\frac{p^i(\mathbf{y}^i|H_1)}{p^i(\mathbf{y}^i|H_0)}.$ (see \cite{nguyen2008dba, tsitsiklis1993decentralized})

 A threshold Bayesian decision rule is expressed in terms of the likelihood ratio and is given by
 \begin{equation}\label{DR}
\delta^i_B(\mathbf{y}^i)=\left\{
\begin{array}{cc}
1 & \textrm{~if~} L^i(\mathbf{y}^i)\geq \tau^i  \\
 0 &    \textrm{~if~} L^i(\mathbf{y}^i)< \tau^i    
\end{array}
 \right.,
 \end{equation}
 where the threshold $\tau^i$ is defined by
\begin{equation}\label{tau}
 \tau^i=\frac{(C^i_{10}-C^i_{00})\pi^i_0}{(C^i_{01}-C^i_{11})\pi^i_1}.
 \end{equation}

If the costs are symmetric and the two hypothesis are equal likely, then the  rule in (\ref{DR}) reduces to the  maximum likelihood (ML) decision rule
  \begin{equation}\label{MLDR}
\delta^i_{ML}(\mathbf{y})=\left\{
\begin{array}{cc}
1  & \textrm{~if~} p^i(\mathbf{y}^i|H_1)\geq    p^i(\mathbf{y}^i|H_0)  \\
 0 &    \textrm{~if~} p^i(\mathbf{y}^i|H_1)< p^i(\mathbf{y}^i|H_0)  
\end{array}
 \right.,
 \end{equation}
 \vspace{-3mm}
 \subsection{Sequential Hypothesis Testing}
 In this section, we use sequential hypothesis testing to make decisions with minimum number of feedback from the peer IDSs, \cite{Wald47}, \cite{Levy08}. An IDS asks for feedback from its acquaintance list until a sufficient number of answers are collected. Let $\Omega^i$ denote all the possible collections of feedback in the acquaintance list to an IDS $i$ and $\omega^i\in\Omega^i$ denotes a particular collection of feedback. Let $N^i(\omega^i)$ be a random variable denoting the number of feedbacks used until a decision is made. A sequential decision rule is formed by a pair $(\phi, \delta)$, where $\phi^i=\{\phi_n^i, n\in\mathbb{N}\}$ is a stopping rule and $\delta^i=\{\delta_n^i, n\in\mathbb{N}\}$ is the terminal decision rule. Introduce a stopping rule with $n$ feedback, $\phi^i_n: \mathcal{Y}_n^i:=\prod_{j\in \mathcal{N}_{i,n}}\mathcal{Y}_j \rightarrow \{0,1\}$, where $\mathcal{N}_{i,n}$ is the set of nodes an IDS $i$ asks up to time $n$. $\phi^i_n=0$ indicates that IDS $i$ needs to take more samples after $n$ rounds whereas $\phi^i_n=1$ means to stop asking for feedback and a decision can be made by the rule $\delta_n^i$. The minimum number of feedbacks is given by
 \begin{equation}
 N^i(\omega^i)=\min\{n:\phi_n^i=1, n\in\mathbb{N}\}.
 \end{equation}
 Note that $N^i(\omega^i)$ is the stopping time of the decision rule. The decision rule $\delta^i$ is not used until $N.$
 We assume that no cost has incurred when a correct decision is made while the cost of a missed intrusion is denoted by $C^i_M$ and the cost of a false alarm is denoted by $C_F^i$. In addition, we assume each feedback incurs a cost $D^i$. We introduce an optimal sequential rule that minimizes Bayes risk given by
 \begin{equation}\label{BR}
 R^i(\phi^i,\delta^i)=R(\phi^i,\delta^i|H_0)\pi_0^i+R(\phi^i,\delta^i|H_1)\pi_1^i,
 \end{equation}where $R(\phi^i,\delta^i|H_l), l=0,1$, are the Bayes risks under hypotheses $H_0$ and $H_1$, respectively:
 {\small \begin{eqnarray}
\nonumber R^i(\phi^i,\delta^i|H_0)=C_F^i\mathbb{P}[\delta_N(Y_j,j\in\mathcal{N}_{i,N})=1|H_0]+ D^i\mathbb{E}[N|H_0],\\
\nonumber  R^i(\phi^i,\delta^i|H_1)=C_M^i\mathbb{P}[\delta_N(Y_j,j\in\mathcal{N}_{i,N})=0|H_1]+ D^i\mathbb{E}[N|H_1].
 \end{eqnarray}}
 Let $ V^i(\pi_0^i)=\min_{\phi^i,\delta^i}R^i(\phi^i,\delta^i)$ be the optimal value function. It is clear that when no feedback are obtained from the peers, the Bayes risks reduce to
 \begin{eqnarray}
 R^i(\phi_0^i=1, \delta_0^i=1)&=&C_F^i\pi_0^i,\\
 R^i(\phi_0^i=1,\delta_0^i=0)&=&C_M^i\pi_1^i.
 \end{eqnarray}
 Hence, $H_1$ is chosen when $C_F^i\pi_0^i<C_M^i\pi_1^i$ or $\pi_0<\frac{C_M^i}{C_F^i+C_M^i}$, and $H_0$ is chosen otherwise. The minimum Bayes risk under no feedback is thus obtained as a function of $\pi_0^i$ and is denoted by
 \begin{equation}\label{TerminalRule}
 T^i(\pi_0^i)=\left\{
 \begin{array}{ll}
 C_F^i\pi_0^i &\textrm{~if~}\pi_0<\frac{C_M^i}{C_F^i+C_M^i},\\
C^i_M(1-\pi_0^i) & \textrm{~otherwise.~}
 \end{array}\right.
 \end{equation}
 The minimum cost function (\ref{TerminalRule}) is a piecewise linear function.
For $\phi^i$ such that $\phi_0^i=0$, i.e., at least one feedback is obtained, let the minimum Bayes risk be denoted by $J^i(\pi_0^i)=\min_{\{(\phi^i,\delta^i): \phi^i_0=0\}}R^i(\phi^i,\delta^i)$. Hence, the optimal Bayes risk needs to satisfy 
 \begin{equation}\label{Vi}
 V^i(\pi_0^i)=\min\{T^i(\pi_0^i), J^i(\pi_0^i)\}.
 \end{equation} 
 Note that $J^i(\pi_0^i)$ must be greater than the cost of one sample $D^i$ as a sample request incurs $D^i$ and $J^i(\pi_0^i)$ is concave in $\pi_0^i$ as a consequence of minimizing the linear Bayes risk (\ref{BR}).
 If the cost $D^i$ is high enough so that $J^i(\pi_0^i)>T^i(\pi_0^i)$ for all $\pi_0^i$, then no feedback will be requested. In this case, $V^i(\pi_0^i)=T^i(\pi_0^i),$ and the terminal rule is described in (\ref{TerminalRule}). For other values of $D^i>0$, due to the piecewise linearity of $T^i(\pi_0^i)$ and
concavity of $J^i(\pi_0^i)$, we can see that $J^i(\pi_0^i)$ and $T^i(\pi_0^i)$ have two intersection points $\pi_L^i$ and $\pi_H^i$ such that $\pi_L^i\leq\pi_H^i$. It can be shown that for some reasonably low cost $D^i$ and $\pi_0^i$ such that $\pi_L^i<\pi_0^i<\pi_H^i$, an IDS optimizes its risk by requesting another feedback; otherwise, an IDS should choose to raise an alarm when $\pi_0^i\leq \pi^i_L$ and report no intrusion when $\pi_0^i\leq \pi^i_L$. 
 
 Assuming that it takes the same cost $D^i$ for IDS $i$ to acquire a feedback, the problem has the same form after obtaining a feedback from a peer. IDS $i$ can use the feedback to update its apriori probability. After $n$ feedback are obtained, $\pi_{0}^i$ can be updated as follows:
 \begin{eqnarray}
\pi_{0}^i(n)
 &=&\frac{\pi_0^i}{\pi_0^i+(1-\pi_0^i)L^i_n};
 \end{eqnarray}
 where $L_n^i:=\prod_{j\in\mathcal{N}_{i,n}}\frac{p(y_j|H_1)}{p(y_j|H_0)}.$
We can thus obtain the optimum Bayesian rule captured by Algorithm 1 below, known as the sequential probability ratio test (SPRT) for a reasonable cost $D^i$. 

\begin{algorithm}
\caption{SPRT Rule for an IDS $i$}          
\label{alg1}                           
\textbf{Step 1}: Start with $n=0$. Use (\ref{RULE1}) as a stopping rule until $\phi_n^i=1$ for some $n\geq0$. \begin{equation}\label{RULE1}
\phi_n^i=\left\{
 \begin{array}{ll}
0 &\textrm{~if~}\pi_L^i<\pi_0^i(n)<\pi_H^i,\\
1 & \textrm{~otherwise.~}
 \end{array}\right.
\end{equation} or in terms of the likelihood ratio $L_n^i$, we can use $
\phi_n^i=\left\{
 \begin{array}{ll}
0 &\textrm{~if~}A^i<L_n^i<B^i\\
1 & \textrm{~otherwise~}
 \end{array}\right.,
$ where  $A^i=\frac{\pi_0^i(1-\pi_H^i)}{(1-\pi_0^i)\pi_H^i}$ and $B^i=\frac{\pi_0^i(1-\pi_L^i)}{(1-\pi_0^i)\pi_L^i}$.
\\
\textbf{Step 2}:  Go to Step 3 if $\phi_n^i=1$ or $n=|\mathcal{N}_i|$; otherwise, choose a new peer from the acquaintance list to request a diagnosis and go to Step 2 with $n=n+1$. \\ 
   \\
\textbf{Step 3}: Apply the terminal decision rule as follows to determine whether there is an intrusion. \\
  \\
  $
    \delta_n^i=\left\{
 \begin{array}{ll}
1 &\textrm{~if~}\pi_0^i(n)\leq\pi_L^i\\
0 & \textrm{~if~}\pi_0^i(n)>\pi_H^i
 \end{array}\right.
  $ or $
    \delta_n^i=\left\{
 \begin{array}{ll}
1 &\textrm{~if~}L_n^i\leq A^i\\
0 & \textrm{~if~}L_n^i> B^i
 \end{array}\right.$
 \\
\end{algorithm}\vspace{-3mm}
%

\vspace{-3mm}
\subsection{Prior Probabilities}
In the above section, the conditional probabilities $p^i(y_i|H_l), i\in\mathcal{N}, l=\{0,1\}$ are assumed to be known. In this section, we use the beta distribution and its Gaussian approximation to find the probabilities. We let $
p^i(y_i=0|H_1):=p^i_M$ be the probability of miss of an IDS $i$'s diagnosis, also known as the false negative (FN) rate; and let $p^i_F:=p^i(y_i=1|H_0)$ be the probability of false alarm or false positive (FP) rate. The probability of detection, or true positive (TP) rate, can be expressed as $p^i_D=1-p^i_M$.

Based on historical data, an IDS $j$ can assess the distributions over its peer IDS $i$'s probabilities of detection and false alarm as beta functions parameterized by $\alpha_i^F, \alpha_i^D$ and $\beta_i^F, \beta_i^F$;{\small
\begin{eqnarray}\label{p}
 p^i_F \sim &\textrm{Beta}(x^i|\alpha^i_F,\beta^i_F)=\frac{\Gamma(\alpha^i_F+\beta^i_F)}{\Gamma(\alpha_F^i)\Gamma(\beta^i_F)}x_i^{\alpha^i_F-1}(1-x_i)^{\beta^i_F-1},\\
p^i_D \sim &\textrm{Beta}(y_i|\alpha^i_D,\beta^i_D)=\frac{\Gamma(\alpha^i_D+\beta^i_D)}{\Gamma(\alpha^i_D)\Gamma(\beta^i_D)}y_i^{\alpha^i_D-1}(1-y_i)^{\beta^i_D-1},
\end{eqnarray}}
where $x_i, y_i\in [0, 1]$; $\alpha_i^F, \alpha_i^D$ and $\beta_i^F, \beta_i^F$ are beta function parameters that are updated according to historical data as follows.
{\small
\begin{eqnarray}\label{alphabeta}
 \alpha^i_F=\sum_{k\in\mathcal{M}_0}(\lambda_F^i)^{t_k^i}r^i_{F,k},
 & \beta^i_F=\sum_{k\in\mathcal{M}_0}(\lambda_F^i)^{t_k^i}(1-r^i_{F,k});\\
\alpha^i_D=\sum_{k\in\mathcal{M}_1}(\lambda_D^i)^{t_k^i}r^i_{D,k},
 & \beta^i_D=\sum_{k\in\mathcal{M}_1}(\lambda_D^i)^{t_k^i}(1-r^i_{D,k}).
\end{eqnarray}
}
The introduction of the discount factors $\lambda_F^i, \lambda_D^i\in[0,1]$ allows more weights on recent data from IDS $i$ while less on the old ones. The discount factors on the data can be different for false negative and false positive rates. The parameter $t_k^i$ denotes the time when $k$-th diagnosis data is generated (and sent to its peer) by IDS $i$. The parameter $r^i_{F,k}, r^i_{M,k}\in[0, 1]$ is the revealed results of the $k$-th diagnosis data: $r^i_{F,k}=1$ suggests that the $k$-th diagnosis data from peer $i$ yields a undetected intrusion while $r^i_{F,k}=0$ means otherwise; similarly, $r^i_{D,k}=1$ indicates the data from the peer $i$ results in a correct detection under intrusion and $r^i_{D,k}=0$ suggests otherwise. The total reported diagnosis data is the set $\mathcal{M}$ and they are classified into two groups: one is where the result is either false positive or true negative under no intrusion, denoted by the set $\mathcal{M}_0$; and the other is where the result is either false negative or true positive under intrusion, denoted by the set $\mathcal{M}_1$. Both sets are disjoint satisfying $\mathcal{M}_0\cup\mathcal{M}_1=\mathcal{M}$ and $\mathcal{M}_0\cap\mathcal{M}_1=\emptyset$.

Each peer $j$ can assess a peer $i$ using (\ref{p}) and (\ref{alphabeta}), where we have not included index $j$ in the expressions for simplicity. However, it is clear that (\ref{p}) and (\ref{alphabeta}) are assessed from the perspective of a certain IDS $j$. In addition, the discount factors in (\ref{alphabeta}) need not be the same for all $j$. Hence, we can implicitly view (\ref{alphabeta}) dependent on $j$.

When parameters of the beta functions $\alpha$ and $\beta$ in (\ref{p}) are sufficiently large, i.e., enough data are collected, beta distribution can be approximated by a Gaussian distribution as
\begin{equation}\label{GApproxExp}
\textrm{Beta}(\alpha, \beta) \approx N\left(\frac{\alpha}{\alpha+\beta},\sqrt{\frac{\alpha\beta}{(\alpha+\beta)^2(\alpha+\beta+1)}}\right).
\end{equation}
Note that we have dropped the superscripts and subscripts in (\ref{GApproxExp}) for generality as it can be applied to all $i$ in (\ref{p}).
 Hence, using the Gaussian approximation and (\ref{alphabeta}), the expected $p^i_D$ and $p^i_M$ are given by
\begin{equation}\label{GApprox}
\mathbb{E}[p^i_F]=\frac{\alpha_F^i}{\alpha_F^i+\beta_F^i},~~
\mathbb{E}[p^i_D]=\frac{\alpha_D^i}{\alpha_D^i+\beta_D^i}.
\end{equation}
The mean values in (\ref{GApprox}) under large data can be intuitively interpreted as the proportion of results of false alarm and detection in the set $\mathcal{M}_0$ and $\mathcal{M}_1$, respectively. They can thus be used in (\ref{condProb}) as the assessment of the peer probability distribution $p_j$. 
\vspace{-3mm}
\subsection{Threshold Approximation}
In the likelihood sequential ratio test of Algorithm \ref{alg1}, the threshold values $A$ and $B$ need to be calculated by finding $\pi^i_L$ and $\pi^i_H$ from $J^i(\pi_0^i)$ and $T^i(\pi_0^i)$ in (\ref{Vi}). The search for these values can be quite involved using dynamic programming. However, in this subsection, we introduce an approximation method to find the thresholds. The approximation is based on theoretical studies made in \cite{Wald47} and \cite{Levy08} where a random walk or martingale model is used to yield a relation between thresholds and false positive and false negative rates. Let $P^i_D, P^i_F$ be the probability of detection and the probability of false alarm of an IDS $i$ after applying the sequential hypothesis testing for feedback aggregation. We need to point out that these probabilities are different from the probabilities $p^i_D, p^i_F$ discussed in the previous subsection, which are the raw detection probabilities without feedback in the collaborative network. Let $\bar{P}_D^i$ and $\bar{P}_F^i$ be reasonable desired performance bounds such that
$
{P}_F^i\leq\bar{P}_F^i,~
{P}_D^i\geq\bar{P}_D^i.
$
Then, the thresholds can be chosen such that
$
A^i=\frac{1-\bar{P}_D^i}{1-\bar{P}_F}^i,~
B^i=\frac{\bar{P}_D^i}{\bar{P}_F^i}.
$

The next proposition gives a result on the bound of the users that need to be on the acquaintance list to achieve the desired performances.
\begin{prop}
Assume that each IDS makes independent diagnosis on their peers' requests and each has the same distribution $p^i_0=\bar{p}_0:=\bar{p}(\cdot| H_0), p^i_1=\bar{p}_1:=\bar{p}(\cdot| H_1)$, $\bar{p}_0(y_i=0)=\theta_0, \bar{p}_1(y_i=0)=\theta_1$, for all $i\in\mathcal{N}$. 

Let $D_{KL}(\bar{p}_0||\bar{p}_1)$ be the Kullback-Leibler (KL) divergence defined as follows.
\begin{eqnarray}
D_{KL}(\bar{p}_0||\bar{p}_1)&=&\sum_{k=0}^1\bar{p}_0(k)\ln\frac{\bar{p}_0(k)}{\bar{p}_1(k)},\\
&=&\theta_0\ln\frac{\theta_0}{\theta_1}+(1-\theta_0)\ln\frac{1-\theta_0}{1-\theta_1}
\end{eqnarray}
Likewise, the K-L divergence $D_{KL}(\bar{p}_1||\bar{p}_0)$ can be defined. 
On average, an IDS needs $N_i$ acquaintances such that
\begin{equation}\label{result}
N_i\geq\max\left(\lceil-\frac{D_M^i}{D_{KL}(\bar{p}_0||\bar{p}_1)}\rceil,\lceil\frac{D_F^i}{D_{KL}(\bar{p}_1||\bar{p}_0)}\rceil\right),
\end{equation}
where $D_M^i=P_F\ln\left(\frac{P_D^i}{P_F^i}\right)+P_D\ln\left(\frac{1-P_D^i}{1-P_F^i}\right)$ and $D_F^i=P_F^i\ln\left(\frac{1-P_D^i}{1-P_F^i}\right)+P_D^i\ln\left(\frac{P_D^i}{P_F^i}\right)$. If $P_F^i\ll 1$ and $P_M^i\ll 1$, we need approximately $N_i$ such that
\begin{equation}\label{approxResult}
N_i\geq\max\left(\lceil\frac{P_D^i-1}{D_{KL}(\bar{p}_0||\bar{p}_1)}\rceil,  \lceil-\frac{P_F^i}{D_{KL}(\bar{p}_1||\bar{p}_0)}\rceil\right).
\end{equation}
\qed
\end{prop}

\begin{proof}
The conditional expected number of feedback needed to reach a decision on the hypothesis in SPRT can be expressed in terms of $P_F$ and $P_D$, \cite{Wald47}, \cite{Levy08}. {\small
\begin{eqnarray}
\nonumber \mathbb{E}[N|H_0]=&\frac{1}{-D_{KL}(\bar{p}_0||\bar{p}_1)}\left[P_F^i\ln\left(\frac{P_D^i}{P_F^i}\right)+P_D^i\ln\left(\frac{1-P_D^i}{1-P_F^i}\right)\right],\\
\nonumber \mathbb{E}[N|H_1]=&\frac{1}{D_{KL}(\bar{p}_1||\bar{p}_0)}\left[P_F^i\ln\left(\frac{1-P_D^i}{1-P^i_F}\right)+P^i_D\ln\left(\frac{P^i_D}{P^i_F}\right)\right],
\end{eqnarray}
}
Hence, to reach a decision we need to have at least $\max\{\mathbb{E}[N|H_0],\mathbb{E}[N|H_1]\}$ independent acquaintances. Under the assumption that both $P_F$ and $P_M^i$ are much less than $1$, we can further approximate $$\mathbb{E}[N|H_0]\sim -\frac{1-P^i_D}{D_{KL}(\bar{p}_0||\bar{p}_1)},\mathbb{E}[N|H_1]\sim -\frac{P^i_F}{D_{KL}(\bar{p}_1||\bar{p}_0)}.$$ These lead us to inequalities (\ref{approxResult}) and (\ref{result}).
\end{proof}
\section{Experiments and Results}
In this section, we use simulations to evaluate the efficiency of the preceding feedback aggregation scheme and compare it with other heuristic approaches, such as the simple average aggregation and the weighted average aggregation. We validate and confirm our theoretical results on the number of acquaintances needed for consultation. The results presented in this section are produced by averaging a large number of replications with negligible confidence intervals. The parameters we use are shown in Table I.

\vspace{-3mm}
\subsection{Simulation Setup}

\begin{table}[t]
\caption{Experimental parameters}
\begin{center}
\begin{tabular}{p{0.9cm}p{0.4cm}p{6.5cm}}
\toprule
  {\small Parameter} & {\small Value} & {\small meaning} \\
\midrule 
  $\tau_{SA}$ & {\small 0.5} &   {\small decision threshold of the simple average model}\\
  $\tau_{WA}$ & {\small 0.5} & {\small decision threshold of the weighted average model}\\
  $d$ & {\small 0.5} & {\small difficulty levels of intrusions and test messages}\\
  $\lambda_F,\lambda_D$ & {\small 0.9} & {\small discount factors in (\ref{alphabeta})}\\
  $\pi_0,\pi_1$ & {\small 0.5} & {\small probability of no-attack and under-attack}\\
  $C_{00},C_{11}$ & {\small 0} & {\small cost of correct decisions}\\
\bottomrule
\end{tabular}
\end{center}
\vspace{-3mm}
\label{esetting}
\end{table}

The simulation environment uses an IDN of $N$ nodes. Each IDS is represented by two parameters, expertise level $l$ and
decision threshold $\tau_p$. At the beginning, each peer receives an initial acquaintance list containing all the other neighbor nodes. In the process of the collaborative intrusion detection, a node sends out requests to its acquaintances for intrusion assessments. The feedback collected are used to make a final decision, i.e., whether to raise an alarm or not. We implement three different feedback mechanisms, namely, simple average aggregation, weighted average aggregation, and hypothesis testing aggregation. We compare their efficiency by the average cost of false decisions.

\subsubsection{Simple Average Model}
If the average of all feedback exceeds a threshold $\tau_{SA}$, then an alarm is raised.
$\tau_{SA}$ is set to $0.5$ if no cost difference is considered for making FP and FN decisions. The simple average mechanism to aggregate feedback is adopted in the literature such as \cite{resnick2006value}.
\subsubsection{Weighed Average Model}
Weights are assigned to feedback from different IDSs to calculate weighted average. Weighted average is widely used to aggregate feedback, such as~\cite{Duma06} and~\cite{Fung2009}, where weights are the trust values of IDSs and trust values are calculated based on their past history. If the weighted average is greater than a threshold $\tau_{WA}$, then an alarm is raised. $\tau_{WA}$ is fixed to $0.5$ in our experiments because their models do not consider the cost difference between FP and FN. In this simulation, we adopt trust values from \cite{Fung2009} as the weights of feedback.

\begin{figure*}[t]
\begin{center}
\begin{minipage}[b]{0.30\linewidth}
\centerline{\psfig{figure=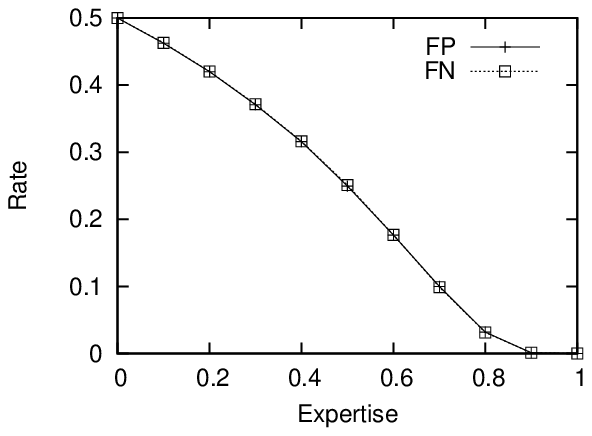,
scale=1.0}} \caption{FP and FN vs. Expertise Level} \label{fig:exp_FPTP}
\end{minipage}
\hspace{0.3cm}
\begin{minipage}[b]{0.30\linewidth} 
\centerline{\psfig{figure=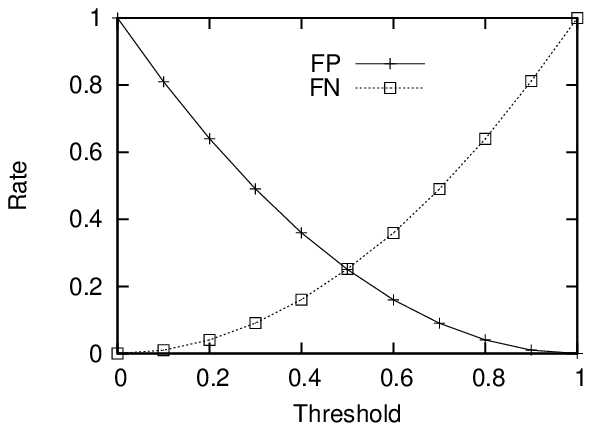,
scale=1.0}} \caption{FP and FN vs. Threshold $\tau_p$} \label{fig:th_FPTP}
\end{minipage}
\hspace{0.3cm}
\begin{minipage}[b]{0.30\linewidth}
\centerline{\psfig{figure=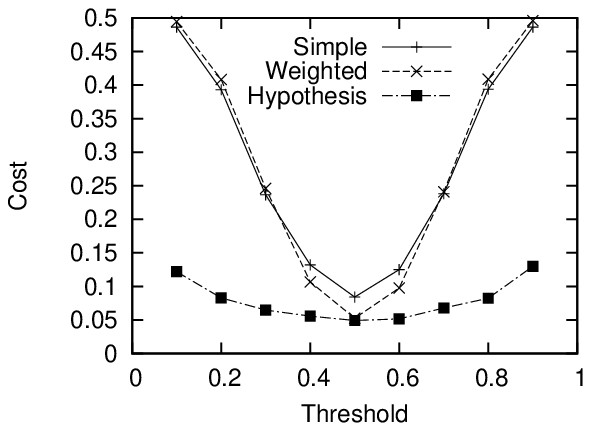,
scale=1.0}} \caption{Average Cost vs. Threshold $\tau_p$}
\label{fig:cost_th_eq}
\end{minipage}
\begin{minipage}[b]{0.30\linewidth}
\centerline{\psfig{figure=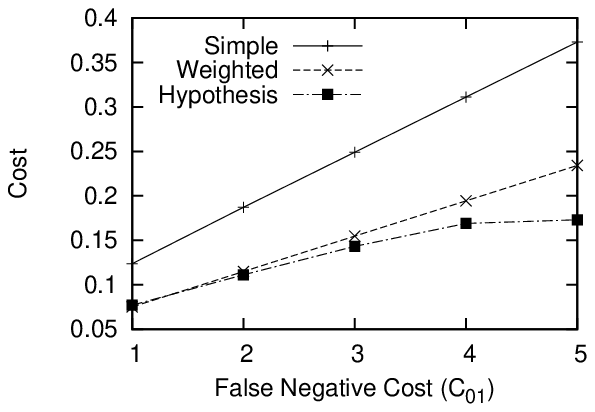,
scale=1.0}} \caption{Cost vs. $C_{01}$ under three models} \label{fig:exp_c01}
\end{minipage}
\hspace{0.3cm}
\begin{minipage}[b]{0.30\linewidth} 
\centerline{\psfig{figure=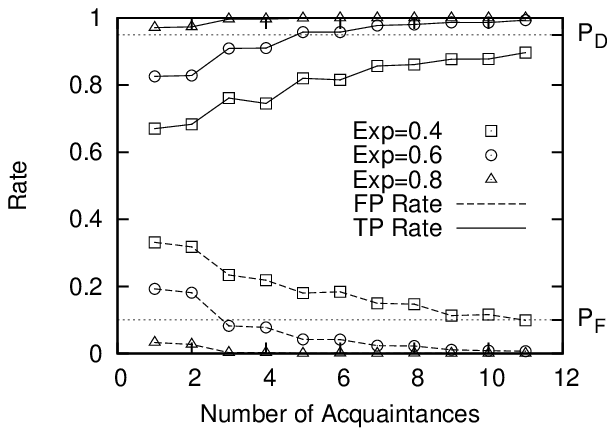,
scale=1.0}} \caption{FP, TP vs. Number of Acquaintances} \label{fig:TF_acq}
\end{minipage}
\hspace{0.3cm}
\begin{minipage}[b]{0.30\linewidth}
\centerline{\psfig{figure=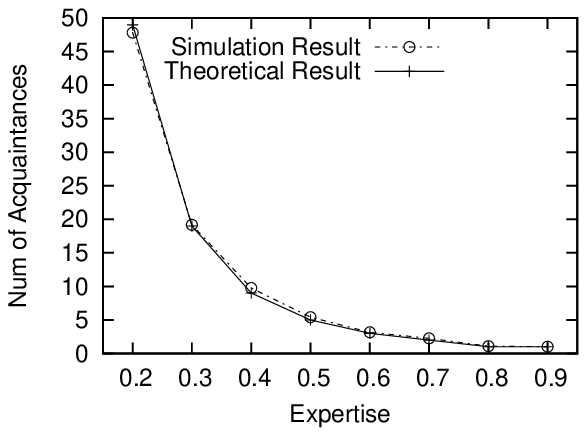,
scale=1.0}} \caption{Number of Acquaintances vs. Expertise}
\label{fig:acq_exp}
\end{minipage}
\end{center}
\end{figure*}
\vspace{-3mm}\subsection{Modeling of an Individual IDS}

To simulate the intrusion detection capability of each node, we use a Beta distribution for the decision model of an IDS. A Beta density function is given by 
\begin{align}
f(\bar{p}|\bar{\alpha},\bar{\beta}) = \frac{1}{B(\bar{\alpha},\bar{\beta})}\bar{p}^{\bar{\alpha}-1}(1-\bar{p})^{\bar{\beta}-1},\\
\nonumber \bar{\alpha} = 1+\frac{l(1-d)}{d(1-l)} r, ~~
\bar{\beta} =1+\frac{l(1-d)}{d(1-l)} (1-r).
\end{align}
where $B(\bar{\alpha},\bar{\beta}) = \int_0^1 t^{\bar{\alpha}-1} (1-t)^{\bar{\beta}-1} dt$, $\bar{p}\in[0,1]$ is the probability of intrusion assessed by the host IDS. $f(\bar{p}|\bar{\alpha},\bar{\beta})$ is the probability that a peer with expertise level $l\in [0,1]$ answers with a value of $\bar{p}$ to an intrusion assessment of difficulty level $d\in [0,1]$. Higher values of $d$ are associated with attacks that are difficult to detect, i.e., many peers may fail to identify them. Higher values of $l$ imply a higher probability of producing correct intrusion assessment. $r\in\{0,1\}$ is the expected result of detection. $r=1$ indicates that there is an intrusion and $r=0$ indicates that there is no intrusion.

Let $\tau_p$ be the decision threshold of $\bar{p}$. If $\bar{p}>\tau_p$, a peer sends feedback $1$ (i.e., under-attack); otherwise, feedback $0$~(i.e., no-attack) is generated.  
%

For a fixed difficulty level, the preceding model assigns higher probabilities of producing correct intrusion diagnosis to peers with higher level of expertise. 
$l=1$ or $d=0$ represent extreme cases where the peer can always accurately detect the intrusion. This is reflected in the Beta distribution with $\bar{\alpha},\bar{\beta}\rightarrow \infty$.

Figure \ref{fig:exp_FPTP} shows that both the FP and FN decrease when the expertise level of an IDS increases. We notice that the curves of FP rate and FN rate overlap. This is because the IDS detection density distributions are symmetric under $r=0$ and $r=1$. Figure \ref{fig:th_FPTP} shows that the FP rate decreases with the decision threshold while the FN rate increases with the decision threshold. When the decision threshold is $0$, all feedback are positive (under-attack); when the decision threshold is $1$, all feedback are negative (no-attack).
\vspace{-3mm}
\subsection{Detection Accuracy and Cost}

One of the most important metrics to evaluate a feedback aggregation scheme is the cost of incorrect decisions. In this experiment, we study the costs of the three aggregation models using a simulated network. We set $N=10$ and fix the expertise level $l$ of all nodes to $0.5$ and set $C_{10}=C_{01}=1$ in (\ref{tau}) for the fairness of comparison, since the simple average and the weighted average models do not account for the cost difference between FP and FN. We fix the decision threshold for each IDS ($\tau_p$) to $0.1$ for the first batch run and then increase it by $0.1$ in each subsequent batch run until it reaches $0.9$. We measure the cost of the three models. As shown in Figure \ref{fig:cost_th_eq}, the costs yielded by the aggregation using hypothesis testing remains the lowest among the three under all threshold settings. The costs of the weighted average and the simple average are close to each other. This is because in this experiment, the weights of all IDSs are the same. Therefore, the difference between the weighted average and the simple average is not substantial. We also observe that changing the threshold has a big impact on the costs of the weighted average and the simple average, while the cost of the hypothesis testing changes only slightly with the thresholds. All costs reach a minimum when the threshold is $0.5$ and increase when it deviates from $0.5$. 

In the next experiment, the expertise levels of all nodes remain $0.5$ and their decision thresholds vary from $0.1$ to $0.9$. We set $C_{10}=C_{01}=1$ in the first batch run and increase $C_{01}$ by $1$ in every subsequent batch run. We observe the costs under three different models. Figure \ref{fig:exp_c01} shows that the costs of the simple average model and the weighted average model increase linearly with $C_{01}$ while cost of hypothesis testing model grows the slowest among the three. This is because the hypothesis testing model has a flexible threshold to optimize its cost. The hypothesis testing model has superiority when the cost difference between FP and FN is large. 
\vspace{-3mm}
\subsection{Sequential Consultation}
In this experiment, we study the number of acquaintances needed for consultation to reach a predefined goal. Suppose the TP lower-bound $\bar{P}_D=0.95$ and FP upper-bound $\bar{P}_F=0.1$. We observe the change of FP rate and TP rate with the number of acquaintances consulted ($n$). Figure \ref{fig:TF_acq} shows that FP rate decreases and TP rate increases with $n$. Consulting higher expertise nodes leads to a higher TP rate and a lower FP rate. In the next experiment we implement Algorithm 1 on each node and measure the average number of acquaintances needed to reach the predefined TP lower-bound and the FP upper-bound. Figure \ref{fig:acq_exp} compares the simulation results with the theoretical results (see (\ref{approxResult})), where the former confirms the latter. In both cases, the number of consultations decreases quickly with the expertise levels of acquaintances. For example, the IDS needs to consult around $50$ acquaintances of expertise $0.2$, while only $3$ acquaintances of expertise $0.7$ are needed for the same purpose. This is partly because low expertise nodes are more likely to make conflicting feedbacks and consequently increase the number of consultations. The analytical results can be useful for IDSs to design the size of their acquaintance lists.
 \section{Conclusion} \label{conclusion}
In this paper, we have presented a sequential hypothesis testing approach to feedback aggregation in a collaborative intrusion detection network. In this mechanism, an IDS consults sequentially for peer diagnoses until it is capable of making an aggregated decision that satisfies Bayes optimal cost criterion. The decision is made based on a threshold rule leveraging the likelihood ratio approximated by beta distribution and thresholds by target rates. Our experimental results show that our proposed feedback aggregation model is superior to other proposed models in the literature in terms of cost efficiency. Our simulation results have also corroborated our theoretical results on the average number of acquaintances needed to reach the predefined false positive upper-bound and true positive lower-bound. As future work, we intend to investigate the robustness of the collaboration system against malicious insiders, especially under collusion attacks. Furthermore, we aim to extend our results to deal with the case of correlated feedbacks.

\bibliographystyle{IEEEtran}
\bibliography{bibfile}
%

%

\end{document}